\documentclass{article}
\usepackage{spconf,amsmath,graphicx,bm,amssymb}
\usepackage{enumitem}
\usepackage{multirow}
\usepackage{url}

\title{Stage-Wise and Prior-Aware Neural Speech Phase Prediction} 
\name{Fei Liu, Yang Ai\sthanks{Corresponding author. This work was funded by the National Nature Science Foundation of China under Grant 62301521, the Anhui Provincial Natural Science Foundation under Grant 2308085QF200, and the Fundamental Research Funds for the Central Universities under Grant WK2100000033.}, Hui-Peng Du, Ye-Xin Lu, Rui-Chen Zheng, Zhen-Hua Ling}
\address{National Engineering Research Center of Speech and Language Information Processing, \\
University of Science and Technology of China, Hefei, P. R. China \\
{\small \tt {fliu215@mail.ustc.edu.cn},
{yangai@ustc.edu.cn}}, \\
{\small \tt{\{redmist,yxlu0102,zhengruichen\}@mail.ustc.edu.cn},
{zhling@ustc.edu.cn}}}
\begin{document}
\ninept
\maketitle

\begin{abstract}

This paper proposes a novel Stage-wise and Prior-aware Neural Speech Phase Prediction (SP-NSPP) model, which predicts the phase spectrum from input amplitude spectrum by two-stage neural networks. In the initial prior-construction stage, we preliminarily predict a rough prior phase spectrum from the amplitude spectrum. The subsequent refinement stage transforms the amplitude spectrum into a refined high-quality phase spectrum conditioned on the prior phase. Networks in both stages use ConvNeXt v2 blocks as the backbone and adopt adversarial training by innovatively introducing a phase spectrum discriminator (PSD). To further improve the continuity of the refined phase, we also incorporate a time-frequency integrated difference (TFID) loss in the refinement stage. Experimental results confirm that, compared to neural network-based no-prior phase prediction methods, the proposed SP-NSPP achieves higher phase prediction accuracy, thanks to introducing the coarse phase priors and diverse training criteria. Compared to iterative phase estimation algorithms, our proposed SP-NSPP does not require multiple rounds of staged iterations, resulting in higher generation efficiency.

\end{abstract}
\vspace{-1mm}
\section{Introduction}
\label{sec:intro}
\vspace{-1mm}
Speech phase prediction is a crucial task in the field of speech signal processing. 
The phase information of speech signals plays a vital role in numerous speech generation tasks, e.g., speech synthesis (SS) \cite{zen2009statistical,takaki2017direct,shen2018natural,marafioti2019adversarial,saito2018text}, speech enhancement (SE) \cite{lu2013speech,xu2014regression,kim2020t}, bandwidth extension (BWE) \cite{wang2015speech,gu2016speech,li2015dnn}, etc. 
Currently, most of the above tasks focus on predicting the amplitude information of speech signals or derived features (e.g., mel spectrograms and mel cepstra). 
Therefore, predicting phase information for these tasks remains to be explored. 
The speech phase prediction aims to recover the missing or unknown phase information from the known amplitude information, thereby restoring the complete short-time spectral information which can be converted to speech waveform via inverse short-time Fourier transform (ISTFT). 

Early research on speech phase prediction primarily focused on iterative estimations of the phase, such as the well-known Griffin-Lim algorithm (GLA) \cite{griffin1984signal}. 
GLA estimates the phase spectrum from the amplitude spectrum by iteratively executing STFT and ISTFT. 
In each iteration (except the first one), GLA uses the phase generated from the previous iteration as a prior, progressively refining the phase.
Its implementation is relatively simple, which has led to its widespread application in various speech generation tasks. 
However, the accuracy of the phase estimated by GLA and some of its variants \cite{perraudin2013fast,masuyama2018griffin} remains unsatisfactory due to their overly simplistic alternating projection operators. 
Recently, Kobayashi \MakeLowercase{\textit{et al.}} \cite{kobayashi2022acoustic} has proposed applying the relaxed averaged alternating reflection (RAAR) algorithm from the optics community to speech phase prediction, utilizing a more complex alternating reflection operator, which has shown impressive results. 
However, the complex iterative operator severely impacts the efficiency of phase estimation.


With the advancement of deep learning, methods combining traditional iterative algorithms and neural networks have emerged.
For example, Masuyama \MakeLowercase{\textit{et al.}} \cite{masuyama2019deep,masuyama2020deep} introduced a deep Griffin-Lim iteration (DeGLI), which utilizes a trainable neural network to simulate the GLA process and achieve iterative phase reconstruction. 
Takamichi \MakeLowercase{\textit{et al.}} \cite{takamichi2018phase,takamichi2020phase} employed a prior-distribution-aware approach, assuming that the phase follows a specific prior distribution (e.g., von Mises distribution or sine-skewed generalized cardioid distribution), and then uses a deep neural network (DNN) to predict the phase information.
However, the phase predicted by the DNN still needs to be refined through GLA iterations. 
Therefore, this type of methods has the disadvantages of high complexity and low efficiency.

\begin{figure*}
\includegraphics[width=0.986\linewidth]{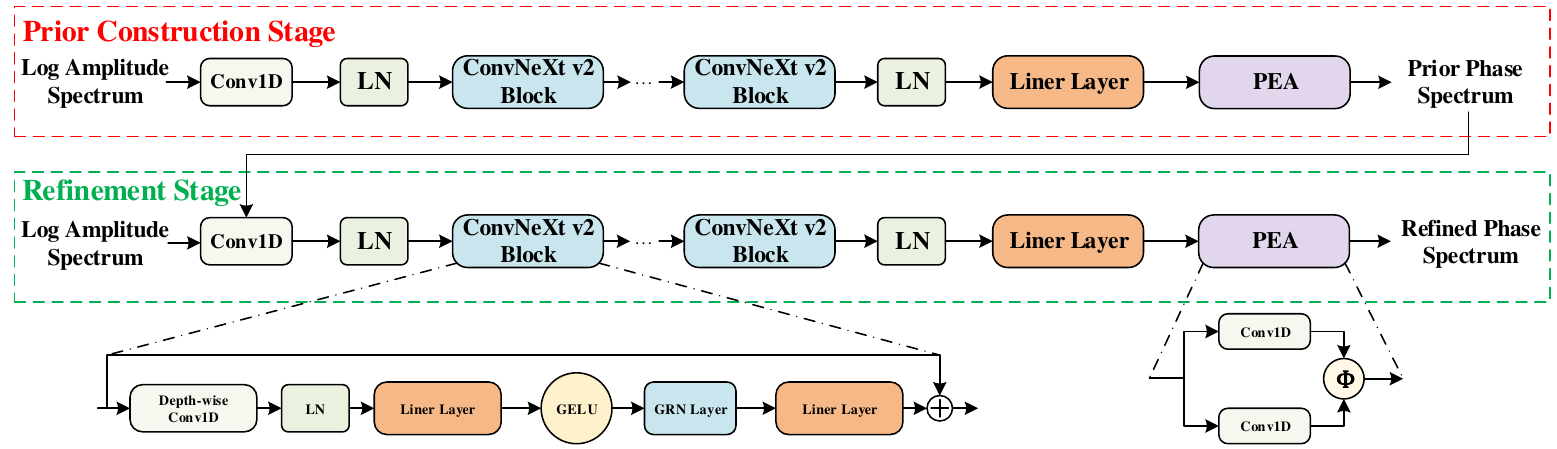}
\centering
\caption{Details of the model structure of the proposed SP-NSPP. Here, \emph{Conv1D}, \emph{LN}, \emph{PEA}, \emph{GELU}, \emph{GRN} and \emph{$\Phi$} represent the 1D convolutional layer, layer normalization layer, parallel estimation architecture, Gaussian error linear unit, global response normalization and phase calculation formula, respectively.}
\label{fig:SP-NSPP}
\end{figure*}

In recent years, to overcome the accuracy and efficiency bottlenecks in speech phase prediction, researchers have attempted to achieve phase prediction solely using neural network-based approaches. 
In our previous work \cite{ai2023neural,ai2024low}, we have proposed a neural speech phase prediction (NSPP) model, achieving direct phase spectrum prediction from amplitude spectrum only through a neural network. 
The NSPP designs specialized structures and losses tailored to the characteristics of the phase. 
It utilizes a residual convolutional network and a parallel estimation architecture (PEA) to propagate the input log amplitude spectrum and directly output the wrapped phase spectrum. 
The PEA is critical to direct phase prediction and consistes of two parallel convolutional layers and a phase calculation formula. 
During training, NSPP proposes an anti-wrapping phase loss, which effectively reduces the error between the predicted and natural phase, which is vital for accurating phase prediction. 
Experimental results have confirmed that NSPP has higher phase prediction accuracy and faster prediction speed than traditional iterative algorithms.

However, NSPP still has some limitations. 
Firstly, NSPP relies solely on amplitude information as input, without prior phase information, making the learning process more challenging.
Secondly, the backbone of NSPP must be updated, as it is complex and redundant and has limited modeling capabilities. 
Lastly, the loss function used by NSPP is too simple, limiting the accuracy of phase prediction.
Therefore, we propose a novel stage-wise and prior-aware NSPP (SP-NSPP) model. 
The core of SP-NSPP lies in incorporating prior phases by referencing iterative algorithms, which significantly enhances the accuracy of phase prediction. 
In the initial prior construction stage, we preliminarily predict a coarsely estimated prior phase spectrum from the amplitude spectrum. 
The subsequent refinement stage converts the amplitude spectrum into a finely refined high-quality phase spectrum conditioned on the previous phase. 
Both stages employ the same ConvNeXt v2-based backbone network and incorporate adversarial training strategies with a phase spectrum discriminator (PSD).
To further refine the time-frequency continuity of the phase, we also introduce a time-frequency integrated difference (TFID) loss in the refinement stage.
Experimental results confirm that, compared to no-prior NSPP, SP-NSPP achieves significantly higher phase prediction accuracy.
Additionally, compared to some iterative algorithms like GLA and RAAR, SP-NSPP requires no multiple iterations, resulting in higher efficiency.

The organization of this paper is as follows.
Section \ref{sec:propose} presents a detailed description of the proposed SP-NSPP model. 
Section \ref{sec:exp} presents our experimental results.
Finally, we give the conclusion in Section \ref{sec:con}.

\vspace{-1mm}
\section{PROPOSED METHOD}
\label{sec:propose}
\vspace{-1mm}
\subsection{Overview}
\vspace{-1mm}
An overview of the proposed SP-NSPP architecture is shown in Figure \ref{fig:SP-NSPP}. 
The SP-NSPP is a two-stage model that predicts the phase spectrum $\hat{\bm{P}}_{refine}\in\mathbb{R}^{F\times N}$ from the input log amplitude spectrum $\bm{A}\in\mathbb{R}^{F\times N}$, using the prior phase spectrum $\hat{\bm{P}}_{prior}\in\mathbb{R}^{F\times N}$ as a bridge, where $F$ and $N$ denote the number of frames and frequency bins, respectively.

\begin{itemize}[nosep, leftmargin=*]
\item {}{\textbf{Prior Construction Stage}}: 
In this stage, only the log amplitude spectrum $\bm{A}$ is used as input to preliminarily predict a coarse phase spectrum $\hat{\bm{P}}_{prior}$ as prior for subsequent stage, i.e.,
\begin{equation}
\hat{\bm{P}}_{prior}=\text{Model}_{\text{PC}}(\bm{A}),
\end{equation}
where $\text{Model}_{\text{PC}}$ is the prior construction model. 
\item {}{\textbf{Refinement Stage}}: 
In this stage, the refinement model converts the log amplitude spectrum $\bm{A}$ into the final refined phase spectrum $\hat{\bm{P}}_{refine}$, conditioned on the prior phase spectrum $\hat{\bm{P}}_{prior}$, i.e., 
\begin{equation}
\hat{\bm{P}}_{refine}=\text{Model}_{\text{R}}(\bm{A}|\hat{\bm{P}}_{prior}),
\end{equation}
where $\text{Model}_{\text{R}}$ is the refinement model. 
Introducing prior information is expected to enable the refinement model to achieve more accurate phase prediction based on this prior, thereby reducing the learning difficulty compared to models without prior information (e.g., NSPP \cite{ai2023neural,ai2024low}). 
\end{itemize}
Finally, the input log amplitude spectrum $\bm{A}$ and the refined phase spectrum $\hat{\bm{P}}_{refine}$ are used to reconstruct the speech waveform $\hat{\bm{x}}\in\mathbb{R}^T$ through ISTFT, where $T$ denotes the waveform sample numbers.


\vspace{-1mm}
\subsection{Model Structure}
\vspace{-1mm}
As shown in Figure \ref{fig:SP-NSPP}, the prior construction and refinement models share the same structure with different parameters. 
The only difference between the two models is in their inputs. 
For the prior construction model, the input is the log amplitude spectrum. 
For the refinement model, the log amplitude spectrum is concatenated with the conditional prior phase spectrum and used as its input. 


For the prior construction or refinement model, the input first undergoes processing through a 1D convolutional layer. 
It then passes through a layer normalization (LN) \cite{ba2016layer}, followed by deep processing using a ConvNeXt v2 \cite{woo2023convnext} network. 
The output of the ConvNeXt v2 network is further processed through another LN and a liner layer. 
The output of the liner layer is then inputted into the PEA to predict the wrapped phase spectrum.
The ConvNeXt v2 network consists of multiple cascaded ConvNeXt v2 blocks. 
As depicted in Figure \ref{fig:SP-NSPP}, each ConvNeXt v2 block employs a residual connection structure, with the core modules including 1D depth-wise convolutional layer, LN, linear layer, global response normalization (GRN) \cite{woo2023convnext} layer and Gaussian error linear unit (GELU)  activation \cite{hendrycks2016gaussian}.
The PEA is borrowed from the NSPP \cite{ai2023neural,ai2024low}. 
It comprises two parallel 1D convolutional layers and an \emph{atan2} phase calculation formula. 
It mimics the process of calculating the phase spectrum from the real and imaginary parts of the complex spectrum and strictly constrains the wrapped predicted phase values within the principal value range. 
Therefore, PEA is a crucial module for the direct prediction of the wrapped phase.
\vspace{-1mm}
\subsection{Training Criteria}
\vspace{-1mm}
The training of the prior construction model and the refinement model is hierarchical and separated. 
After the prior reconstruction model is fully trained, it is switched to generation mode to provide data for training the refinement model. 
The training criteria for both models are similar, with the only difference being that the refinement model includes an additional TFID loss.

\begin{figure}
\includegraphics[width=\linewidth]{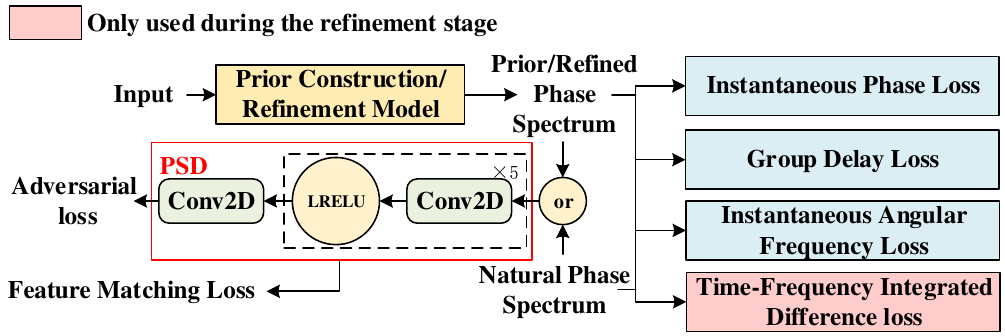}
\caption{Details of the training losses of the proposed SP-NSPP. Here, \emph{Conv2D} and \emph{LRELU} represent the 2D convolutional layer and leaky rectified linear unit, respectively.}
\label{fig:loss}
\end{figure}
\vspace{-1mm}
\subsubsection{Training Criteria of Prior Construction Model}
\label{subsec:PRS}
\vspace{-1mm}
As shown in Figure \ref{fig:loss}, the anti-wrapping losses borrowed from NSPP \cite{ai2023neural,ai2024low} and the newly proposed phase adversarial loss are used to jointly train the prior reconstruction model. 
The anti-wrapping losses are defined between the prior phase spectrum $\hat{\bm{P}}_{prior}$ and the natural one $\bm{P}$ and includes instantaneous phase (IP) loss $\mathcal L_{IP}(\hat{\bm{P}}_{prior},\bm{P})$, group delay (GD) loss $\mathcal L_{GD}(\hat{\bm{P}}_{prior},\bm{P})$, and instantaneous angular frequency (IAF) loss $\mathcal L_{IAF}(\hat{\bm{P}}_{prior},\bm{P})$, i.e.,
\begin{equation}
\resizebox{\linewidth}{!}{
$\mathcal L_{P}(\hat{\bm{P}}_{prior},\bm{P})=\mathcal L_{IP}(\hat{\bm{P}}_{prior},\bm{P})+\mathcal L_{GD}(\hat{\bm{P}}_{prior},\bm{P})+\mathcal L_{IAF}(\hat{\bm{P}}_{prior},\bm{P}).$
}
\end{equation}
These three losses are computed by using an anti-wrapping function $f_{AW}(x)=\left| x-2\pi\cdot round\left( \frac{x}{2\pi} \right) \right|$ to 
activate the direct errors of IP, GD, and IAF, respectively. 
This anti-wrapping function can effectively prevent the issue of training error expansion caused by the phase wrapping characteristics.


For the phase adversarial training, the proposed SP-NSPP incorporates a PSD to ensure high-quality phase prediction. 
As shown in Figure \ref{fig:loss}, the PSD takes either $\hat{\bm{P}}_{prior}$ or $\bm{P}$ as input. 
It consists of five 2D convolutional layers interleaved with leaky rectified linear unit (LReLU) activation to capture time-frequency features. 
The processed features are finally passed through a 2D convolutional layer to output the discriminative values.
During training, the PSD is trained to classify natural phase samples as 1 and generated samples from the generator as 0. 
Conversely, the prior construction model (i.e., the generator) is trained to generate samples that resemble those classified as 1 by the PSD as closely as possible.
We use the adversarial loss with hinge form which is defined as:
\begin{equation}
\label{equ: adv G}
\mathcal L_{adv-G}(\hat{\bm{P}}_{prior},\bm{P})=\mathbb{E}_{\hat{\bm{P}}_{prior}}\max \left(0,1-\text{PSD}(\hat{\bm{P}}_{prior})\right),
\end{equation}
\begin{equation}
\label{equ: adv D}
\begin{aligned}
\mathcal L_{adv-D}(\hat{\bm{P}}_{prior},\bm{P})=\mathbb{E}_{\left(\hat{\bm{P}}_{prior},\bm{P}\right)}\max \left(0,1-\text{PSD}(\bm{P})\right)+ \\ \mathbb{E}_{\left(\hat{\bm{P}}_{prior},\bm{P}\right)} \max \left(0,1+\text{PSD}(\hat{\bm{P}}_{prior})\right).
\end{aligned}
\end{equation}
We also introduce the commonly used feature matching (FM) loss $\mathcal L_{FM}(\hat{\bm{P}}_{prior},\bm{P})$ in vocoder tasks \cite{kong2020hifi,ai2020neural}, defined as the sum of the mean squared errors (MSEs) of the intermediate layer outputs of PSD when taking $\hat{\bm{P}}_{prior}$ or $\bm{P}$ as input.

Therefore, the final loss for the prior construction model is as follows.
\begin{equation}
\label{L G}
\begin{aligned}
\mathcal L(\hat{\bm{P}}_{prior},\bm{P})=\lambda_{P} \mathcal L_{P}(\hat{\bm{P}}_{prior},\bm{P}) +\\ \lambda_{PSD} \left(\mathcal L_{adv-G}(\hat{\bm{P}}_{prior},\bm{P})+\mathcal L_{FM}(\hat{\bm{P}}_{prior},\bm{P}) \right),
\end{aligned}
\end{equation}
where $\lambda_{P}$ and $\lambda_{PSD}$ are hyperparameters. 
The prior construction model and the PSD are trained in an alternating manner with $\mathcal L(\hat{\bm{P}}_{prior},\bm{P})$ and $\mathcal L_{adv-D}(\hat{\bm{P}}_{prior},\bm{P})$, respectively.
\vspace{-1mm}
\subsubsection{Training Criteria of Refinement Model}
\vspace{-1mm}
At the refinement stage, we introduce an additional TFID loss to train the refinement model, compared to training the prior construction model. 
The TFID loss simultaneously considers the differential values of the phase spectrum in both time and frequency directions, further enhancing the temporal and frequency continuity of the phase spectrum for refined optimization.

Given a matrix $\bm{X}\in\mathbb R^{F\times N}$, We first define a series of vector transformation operations within $\bm{X}$ as follows,
\begin{equation}
\label{equ:T}
\Theta_{CL}\bm{X}=\left [ \bm{\omega}_{2}, \bm{\omega}_{3}, \dots, \bm{\omega}_{N}, \bm{0} \right ],
\end{equation}
\begin{equation}
\label{equ:subT}
\Theta_{CR}\bm{X}=\left [\bm{0}, \bm{\omega}_{1}, \bm{\omega}_{2}, \dots, \bm{\omega}_{N-1} \right ], 
\end{equation}
\begin{equation}
\label{equ:F}
\Theta_{RU}\bm{X}=\left [ \bm{\nu}_{2}^\top,  \bm{\nu}_{3}^\top, \dots, \bm{\nu}_{F}^\top, \bm{0}^\top \right ]^\top,
\end{equation}
where $\bm{\omega}_{n}$ and $\bm{\nu}_{f}$ are the $n$-th column vector and $f$-th row vector of matrix $\bm{X}$, respectively. 
Based on this, we define the time-frequency in-direction difference operator $\Delta_{TFIDD}$ and the time-frequency reverse-direction difference operator $\Delta_{TFRDD}$ as follows.
\begin{equation}
\Delta_{TFIDD}\bm{X}=\bm{X}-\Theta_{CL}\Theta_{RU}\bm{X},
\end{equation}
\begin{equation}
\Delta_{TFRDD}\bm{X}=\bm{X}-\Theta_{CR}\Theta_{RU}\bm{X},
\end{equation}

Our proposed TFID loss $\mathcal L_{TFID}(\hat{\bm{P}}_{refine},\bm{P})$ is defined between the refinement phase spectrum $\hat{\bm{P}}_{refine}$ and natural one $\bm{P}$ and includes both time-frequency in-direction difference loss $\mathcal L_{TFIDD}(\hat{\bm{P}}_{refine},\bm{P})$ and time-frequency reverse-direction difference loss $\mathcal L_{TFRDD}(\hat{\bm{P}}_{refine},\bm{P})$, i.e.,
\begin{equation}
\resizebox{\linewidth}{!}{
$\mathcal L_{TFID}(\hat{\bm{P}}_{refine},\bm{P})=\mathcal L_{TFIDD}(\hat{\bm{P}}_{refine},\bm{P})+\mathcal L_{TFRDD}(\hat{\bm{P}}_{refine},\bm{P}),$
}
\end{equation}
where
\begin{equation}
\label{equ:tfid}
\resizebox{\linewidth}{!}{
$\mathcal L_{TFIDD}=\mathbb{E}_{\left(\hat{\bm{P}}_{refine},\bm{P}\right)} \left\lVert f_{AW}\left(\Delta_{TFIDD}\hat{\bm{P}}_{refine}-\Delta_{TFIDD}\bm{P} \right)\right\rVert_1,$
}
\end{equation}
and
\begin{equation}
\label{equ:stfid}
\resizebox{\linewidth}{!}{
$\mathcal L_{TFRDD}=\mathbb{E}_{\left(\hat{\bm{P}}_{refine},\bm{P}\right)} \left\lVert f_{AW}\left(\Delta_{TFRDD}\hat{\bm{P}}_{refine}-\Delta_{TFRDD}\bm{P} \right)\right\rVert_1.$
}
\end{equation}

Therefore, in the refinement stage, we alternately train the refinement model and the PSD using losses $\mathcal L(\hat{\bm{P}}_{refine},\bm{P})+\mathcal L_{TFID}(\hat{\bm{P}}_{refine},\bm{P})$ and $\mathcal L_{adv-D}(\hat{\bm{P}}_{refine},\bm{P})$, respectively.

\vspace{-1mm}
\subsection{Optional Iterative Prediction Mode}
\label{sec: interative}
\vspace{-1mm}
In our proposed SP-NSPP, the predicted phase from the first stage is utilized as the prior phase input for the second stage, resulting in a more refined phase prediction. 
This approach is similar to traditional iterative algorithms. 
Therefore, our proposed SP-NSPP can also adopt an iterative prediction mode. 
In SP-NSPP, the refinement stage can be regarded as performing one iteration based on the prior construction stage. 
Assume $\hat{\bm{P}}_{refine}^0=\hat{\bm{P}}_{prior}$, $\hat{\bm{P}}_{refine}^1=\hat{\bm{P}}_{refine}$ and $\text{Model}_{\text{R}}^1=\text{Model}_{\text{R}}$. 
Then, by introducing more identical refinement models $\text{Model}_{\text{R}}^i (i=2,3,\cdots)$, the iterative prediction mode can be executed as follows. 
\begin{equation}
\hat{\bm{P}}_{refine}^i=\text{Model}_{\text{R}}^i(\bm{A}|\hat{\bm{P}}_{refine}^{i-1}), i=1,2,3,\cdots
\end{equation}
However, as the number of iterations increases, the overall model size grows linearly. 
Therefore, phase prediction accuracy and model complexity should be balanced. 
The relevant experimental analysis is shown in Section \ref{sec: discussion}.


\section{Experiments and Results}
\label{sec:exp}

\subsection{Data and Feature Configuration}

In the experiments, we followed \cite{ai2023neural} to use a subset of the VCTK corpus \cite{veaux2016superseded} consisting of 11,572 speech utterances from 28 speakers. 
The original 48 kHz sampled recordings in the VCTK corpus were downsampled to 16 kHz to ensure a fair comparison with other baseline iterative estimation algorithms and prediction models.
The dataset was randomly constructed into a training set (11,012 utterances) and a validation set (560 utterances). 
We then selected a total of 824 speech utterances from one male unseen speaker and one female unseen speaker as the test set. 
When extracting the amplitude and phase spectrum from the natural waveform, we set the window size to 20 ms, the window shift to 5 ms, and the FFT point number to 1024 (i.e., $N=513$).

\subsection{Task Definitions}

We defined two tasks to compare the performance of different phase estimation or prediction methods. 
\begin{itemize}[nosep, leftmargin=*]
\item {}{\textbf{Analysis-Synthesis Task}}: 
In this task, the phase spectrum is predicted from the natural amplitude spectrum extracted from the natural waveform by STFT. 
This task focuses on evaluating phase recovery and reconstruction capabilities.
\item {}{\textbf{Prediction-Synthesis Task}}: 
In this task, the phase spectrum is predicted from the non-natural amplitude spectrum. 
This non-natural amplitude spectrum is predicted by other models, making it more representative of real-world applications. 
For example, in speech bandwidth extension (BWE), we introduced an amplitude extension model inspired by \cite{ai2024low}. 
This model first predicts the high-frequency amplitude spectrum from the low-frequency one extracted from bandwidth-limited speech, and then concatenates them to construct a full-band amplitude spectrum. 
Finally, the corresponding phase spectrum is recovered by phase prediction methods, and the extended speech waveform is reconstructed through ISTFT. 
This task focuses on evaluating the robustness and generalization of the phase prediction methods.
\end{itemize}


\subsection{Model Details}

The descriptions of phase estimation algorithms and prediction methods for comparison are as follows\footnote{Speech samples can be accessed at \url{https://fliu215.github.io/fliu_demo/}.}.

\begin{itemize}[nosep, leftmargin=*]
\item {}{\textbf{GLA}}: The iterative phase estimation algorithm GLA \cite{griffin1984signal} with 100 iterations.

\item {}{\textbf{RAAR}}: The iterative phase estimation algorithm RAAR \cite{kobayashi2022acoustic} with 100 iterations.

\item {}{\textbf{vMDNN}}: The von Mises distribution-based DNN phase prediction method \cite{takamichi2018phase,takamichi2020phase}. 
We reproduced the DNN model and used it to predict the initial phase spectrum from the amplitude spectrum, then refined it by GLA with 100 iterations.


\item {}{\textbf{NSPP}}: The neural speech phase prediction model NSPP \cite{ai2023neural,ai2024low} which predicted the phase spectrum from the amplitude spectrum without the phase prior. 
We reimplemented it using the official open source code\footnote{\url{https://github.com/yangai520/NSPP}.}.

\item {}{\textbf{SP-NSPP}}: The proposed stage-wise and prior-aware neural speech phase prediction model. 
Here, the prior construction model and refinement model shared the same configuration. 
Each model included eight ConvNeXt v2 blocks. 
All 1D convolutions had a kernel size of 7. 
Except for PEA, the channel size of the 1D convolutions in other parts was uniformly set to 256. 
The channel size of the 1D convolutions in PEA was 513 (i.e., equal to $N$). 
The number of nodes in the first linear layer of each ConvNeXt v2 block was 512, while the number of nodes in the linear layers in other parts was 256. 
For PSD, the first five 2D convolutional layers all had 64 channels, with kernel sizes of 7$\times$5, 5$\times$3, 5$\times$3, 3$\times$3 and 3$\times$3, respectively. 
The 2D convolutional layer for the final output of the discriminative value had one channel and a kernel size of 3. 
The hyperparameters of the loss function were set as $\lambda _{P}=100$ and $\lambda _{PSD}=0.1$. 
Each model was trained using the AdamW optimizer with $\beta=0.8$ on a single Nvidia 2080Ti GPU. 
The initial learning rate was set to 0.0002 for each epoch, with a learning rate decay factor of 0.999. 
The models were trained for a total of 3100 epochs, with a batch size of 16. 
The waveform length was truncated to 8000 samples for each training step.
\end{itemize}

\subsection{Evaluation Metrics}

We comprehensively evaluated and compared the phase prediction methods in terms of phase accuracy, speech quality, and efficiency. 

\begin{table*}[ht]
    \centering
    \caption{Phase accuracy, speech quality and efficiency evaluation results on the test set of VCTK corpus for the analysis-synthesis task.}
    \label{tab:main}
    \resizebox{\textwidth}{!}{
    \begin{tabular}{ccccccccccc}
    \hline
        \textbf{}  & $\textbf{PD}_\textbf{IP}$$\downarrow$ & $\textbf{PD}_\textbf{GD}$$\downarrow$ & $\textbf{PD}_\textbf{IAF}$$\downarrow$ & $\textbf{PD}_\textbf{TFID}$$\downarrow$  &\textbf{SNR(dB)$\uparrow$}& \textbf{PESQ$\uparrow$} & \textbf{F0-RMSE(cent)$\downarrow$}& \textbf{MOS}$\uparrow$ & \textbf{RTF}$\downarrow$ & \textbf{Model Size}$\downarrow$ \\ \hline
        \textbf{Natural} & - & - & - & -  &-& - & - & 3.81 $\pm$ 0.064 & - & -  \\ \hline
        \textbf{GLA} & 1.81& 0.46& 0.84& 0.86&3.35& 3.74& 32.5 & 3.77 $\pm$ 0.061 & 0.208 (4.8$\times$) & -   \\ 
        \textbf{RAAR}  &  1.80& \textbf{0.45}& \textbf{0.60}& \textbf{0.56}&4.52& 4.29& 11.0 & 3.76 $\pm$ 0.063 & 0.396 (2.5$\times$) & - \\ 
        \textbf{vMDNN} & 1.79& \textbf{0.45}& 0.82& 0.84&5.09& 4.09& 13.2 & 3.78 $\pm$ 0.064 & 0.208 (4.8$\times$) & - \\ 
        \textbf{NSPP} & 1.75& 0.58& 1.11& 1.13&8.18& 4.20& 11.3 & 3.78 $\pm$ 0.064  & 0.057 (17.5$\times$) & 147M  \\ 
        \textbf{SP-NSPP} & \textbf{1.72}& 0.50& 0.87& 0.90&\textbf{8.88}& \textbf{4.33}& \textbf{10.7}& \textbf{3.80 $\pm$ 0.065} & \textbf{0.029 (34.5$\times$)} & \textbf{41.3M}\\ \hline
    \end{tabular}}
\end{table*}

\begin{table*}[ht]
    \centering
    \caption{Phase accuracy and speech quality evaluation results on the test set of VCTK corpus for the prediction-synthesis task.}
    \label{tab:bwe}
    \begin{tabular}{cccccccccc}
    \hline
        \textbf{} & $\textbf{PD}_\textbf{IP}$$\downarrow$ & $\textbf{PD}_\textbf{GD}$$\downarrow$ & $\textbf{PD}_\textbf{IAF}$$\downarrow$ & $\textbf{PD}_\textbf{TFID}$$\downarrow$ & \textbf{SNR(dB)$\uparrow$}& \textbf{PESQ$\uparrow$} & \textbf{F0-RMSE(cent)$\downarrow$}   \\ \hline
        \textbf{GLA} & 1.82  & 0.52 & 1.19 & 1.21 & 3.27 & 3.46 & 32.6    \\ 
        \textbf{RAAR} & 1.82  & 0.69 & \textbf{1.13} & \textbf{1.15} & 4.37 & 3.93 & 11.0     \\ 
        \textbf{vMDNN} & 1.79 & \textbf{0.52} & 1.18 & 1.20 & 4.99 & 3.75 & 13.2   \\ 
        \textbf{NSPP} & 1.75 & 0.61 & 1.26 & 1.28 & 8.15 & 3.84 & 11.5   \\ 
        \textbf{SP-NSPP} & \textbf{1.72} & 0.56 & 1.16 & 1.18 & \textbf{8.86} & \textbf{3.93} & \textbf{10.6}   \\ \hline
    \end{tabular}
\end{table*}

\begin{itemize}[nosep, leftmargin=*]
\item {}{\textbf{Phase accuracy evaluations}}: 
To evaluate the phase spectrum prediction accuracy, we proposed a series of phase distortion (PD) metrics. 
The PD metrics first evaluate the phase error using the anti-wrapping function $f_{AW}$ and then calculate the distortion in a manner similar to log-spectral distance (LSD), i.e.,
\begin{equation}
\text{PD}_*=\frac{1}{N} \sum_{n=1}^{N} \sqrt{\frac{1}{F} \sum_{f=1}^{F}f_{AW}^{2}(\Delta_* \bm{P}-\Delta_* \hat{\bm{P}} )  },
\end{equation}
where $\bm{P}\in\mathbb R^{F\times N}$ and $\hat{\bm{P}}\in\mathbb R^{F\times N}$ respectively represent the predicted and natural phase spectra. 
* can be replaced with IP, GD, IAF, TFIDD, and TFRDD, where $\Delta_{\text{IP}}$ denotes no operation, and $\Delta_{\text{GD}}$ and $\Delta_{\text{IAF}}$ represent frequency difference and time difference operations, respectively. 
Since both $\text{PD}_{\text{TFIDD}}$ and $\text{PD}_{\text{TFRDD}}$ calculate the phase differential distortion along the time and frequency axes simultaneously, we compute their average as $\text{PD}_{\text{TFID}}$. 

\item {}{\textbf{Speech quality evaluations}}: 
To evaluate the quality of the speech reconstructed from the amplitude spectrum and the predicted phase spectrum, we used several common objective tools, including signal-to-noise ratio (SNR) and perceptual evaluation of speech quality (PESQ) \cite{rix2001perceptual}. 
We also evaluated the F0 distortion by calculating the root MSE between F0s extracted from reconstructed and natural speeches (denoted by F0-RMSE). 
In terms of subjective evaluation, we employed the mean opinion score (MOS) test to assess the naturalness of the reconstructed speech on the Amazon Mechanical Turk\footnote{\url{https://www.mturk.com.}}. 
At least thirty native English-speaking listeners rate twenty reconstructed speech samples and natural speech samples for each method. 
The scoring range was from 1 to 5 with a 0.5 interval.

\item {}{\textbf{Efficiency evaluations}}: 
In order to assess the generation efficiency of different methods, the real-time factor (RTF) was adopted. 
Additionally, we also measured the size of NSPP-based models to evaluate their complexity. 

\end{itemize}

\subsection{Primary Experimental Results}


First, we compared the proposed \textbf{SP-NSPP} with other baselines for both the analysis-synthesis task and prediction-synthesis task. 
Table \ref{tab:main} shows the results of the analysis-synthesis task. 
Regarding the phase accuracy, our proposed \textbf{SP-NSPP} achieved the lowest $\text{PD}_\text{IP}$, but fell behind iterative algorithms (i.e., the \textbf{GLA}, \textbf{RAAR} and \textbf{vMDNN}) in terms of $\text{PD}_\text{GD}$, $\text{PD}_\text{IAF}$, and $\text{PD}_\text{TFID}$ metrics. 
This conclusion is consistent with that in \cite{ai2024low}. 
We can infer that iterative algorithms strive to improve phase continuity, while neural models focus more on directly reducing instantaneous phase error. 
Compared to \textbf{NSPP}, the proposed \textbf{SP-NSPP} shows significant improvements in all phase metrics, indicating that the introduced phase prior information, as well as the improved structure and training criteria, are effective in enhancing phase prediction accuracy. 

\begin{figure}
\centering
\includegraphics[width=\linewidth]{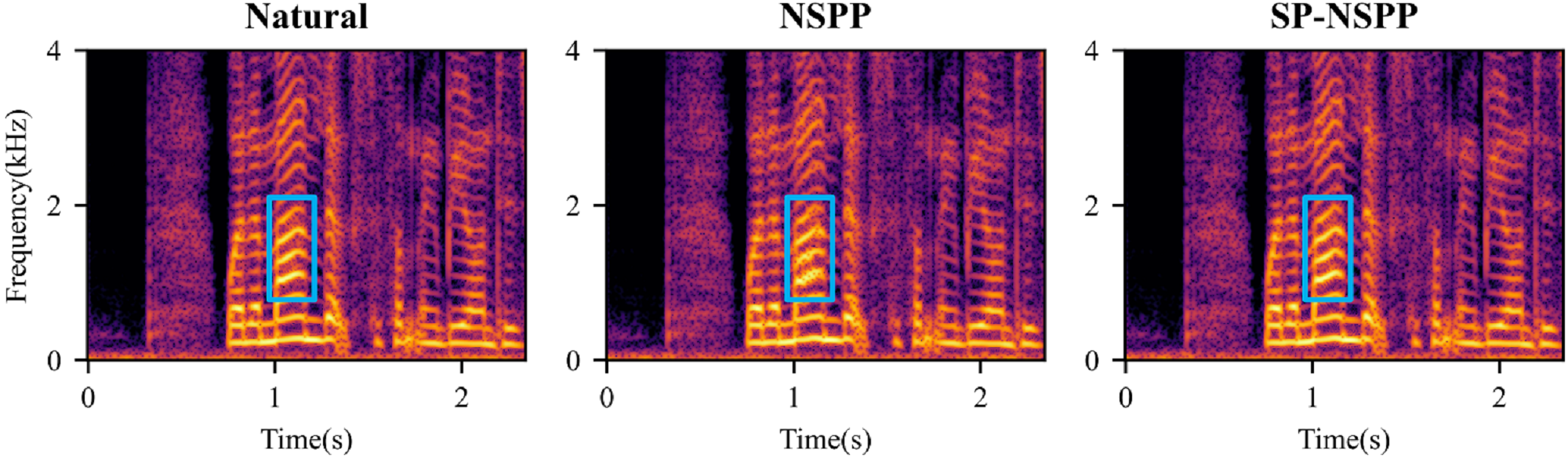}
\caption{A comparison among the spectrograms (0$\sim$4 kHz) of the natural speech and speeches generated by \textbf{NSPP} and \textbf{SP-NSPP} for the analysis-synthesis task.}
\label{fig:spectrum}
\end{figure}

Interestingly, in terms of speech quality, our proposed \textbf{SP-NSPP} obtained the highest SNR, highest PESQ, lowest F0-RMSE and highest MOS score among all phase prediction methods. 
This indicates that the speech reconstructed by \textbf{SP-NSPP} had the best objective and subjective quality. 
We also provided a visual analysis of the spectrograms of natural speech, and the speeches reconstructed by \textbf{NSPP} and \textbf{SP-NSPP} in Figure \ref{fig:spectrum}. 
As shown in the blue box in Figure \ref{fig:spectrum}, the harmonic details in the spectrogram of the speech reconstructed by \textbf{NSPP} are degraded, which is caused by inaccurate phase prediction. 
Because amplitude and phase are coupled, reconstructing the waveform with the natural amplitude spectrum and inaccurately predicted phase spectrum also damages re-extracted amplitude details. 
In contrast, our proposed \textbf{SP-NSPP} is able to restore clear harmonics, thanks to accurate phase prediction. 

The results for the prediction-synthesis task are listed in Table \ref{tab:bwe}. 
The subjective MOS test was excluded from this task. 
We can see that the experimental conclusions for this task are the same as those for the analysis-synthesis task.
Interestingly, the gap between \textbf{SP-NSPP} and iterative algorithms in phase continuity metrics has significantly narrowed compared to the results in the analysis-synthesis task. 
The $\text{PD}_\text{GD}$ of \textbf{SP-NSPP} is even lower than that of \textbf{RAAR}. 
This indicates that our proposed \textbf{SP-NSPP} has better robustness and generalization when using non-natural amplitude spectra as input, making it suitable for application in specific speech generation tasks. 

The experimental results for efficiency evaluation are also shown in Table \ref{tab:main}. 
According to the RTF results, our proposed \textbf{SP-NSPP} had the fastest generation speed. 
Although \textbf{RAAR} is a strong baseline, comparable to \textbf{SP-NSPP} in phase accuracy and speech quality as inferred from Table \ref{tab:main} and \ref{tab:bwe}, its generation speed is only 7.2\% of that of \textbf{SP-NSPP}. 
Though \textbf{SP-NSPP} used two models for two different stages, compared to \textbf{NSPP}'s single model, \textbf{SP-NSPP}'s generation speed is 1.97 times faster, and its model size is only 28.1\% of that of \textbf{NSPP}. 
This indicates that the ConvNeXt v2 backbone network has a smaller model size and higher generation efficiency than the residual convolution network, making it more suitable for phase prediction applications.
Therefore, our proposed \textbf{SP-NSPP} is an efficient and lightweight model with high phase accuracy and high speech quality.

\begin{table*}[!ht]

    \centering
    \caption{phase accuracy and speech quality evaluation results among \textbf{SP-NSPP} and its ablated variants for the analysis-synthesis task.}
    \label{tab:ablation}
    \begin{tabular}{cccccccc}
    \hline
        \textbf{} & $\textbf{PD}_\textbf{IP}$$\downarrow$ & $\textbf{PD}_\textbf{GD}$$\downarrow$ & $\textbf{PD}_\textbf{IAF}$$\downarrow$ & $\textbf{PD}_\textbf{TFID}$$\downarrow$  &\textbf{SNR(dB)$\uparrow$}& \textbf{PESQ$\uparrow$}& \textbf{F0-RMSE(cent)$\downarrow$}  \\ \hline
        \textbf{SP-NSPP} & \textbf{1.72}& \textbf{0.50}& \textbf{0.87}& \textbf{0.90}&8.88& 4.33& 10.7   \\ 
        \textbf{SP-NSPP w/o RS} & 1.75& 0.54& 0.98& 1.01&7.84& 4.23& 11.9  \\ 
        \textbf{SP-NSPP w/o PSD} & 1.73& \textbf{0.50}& \textbf{0.87}& \textbf{0.90}&8.97& 4.34& \textbf{10.3}\\ 
        \textbf{SP-NSPP w/o TFID} & \textbf{1.72}& 0.51& 0.92& 0.94&\textbf{9.16}& \textbf{4.36}& 10.4   \\ 
        \hline
    \end{tabular}
\end{table*}

\begin{table}[!ht]
    \centering
    \caption{Objective evaluation results among phase prediction methods under other data conditions for the analysis-synthesis task.}
    \label{tab:48k}
    \begin{tabular}{ccccc}
    \hline
        \textbf{} & \textbf{} & $\textbf{PD}_\textbf{IP}$$\downarrow$  &\textbf{SNR(dB)$\uparrow$}& \textbf{PESQ$\uparrow$}\\ \hline
        \textbf{RAAR}  & \multirow{3}*{VCTK@24k} & 1.80 &4.64& 3.42\\ 
        \textbf{NSPP} & ~ & 1.71 &11.91& \textbf{4.24}\\ 
        \textbf{SP-NSPP} & ~  & \textbf{1.66} &\textbf{12.92}& 4.19\\ \hline
        \textbf{RAAR} & \multirow{3}*{VCTK@48k}  & 1.81 &5.54& 1.69\\ 
        \textbf{NSPP} & ~  & \textbf{1.73} &\textbf{13.38}& \textbf{3.85}\\ 
        \textbf{SP-NSPP} & ~ & \textbf{1.73} &13.30& 3.72\\ \hline
        \textbf{RAAR} & \multirow{3}*{FSD50K@44.1k} & 1.75 &\textbf{4.33}& -\\ 
        \textbf{NSPP} & ~ & \textbf{1.74} &3.51& -\\ 
        \textbf{SP-NSPP} & ~ & \textbf{1.74} &3.39& -\\ \hline
    \end{tabular}
\end{table}

\subsection{Ablation Studies}

\begin{figure}
\centering
\includegraphics[width=\linewidth]{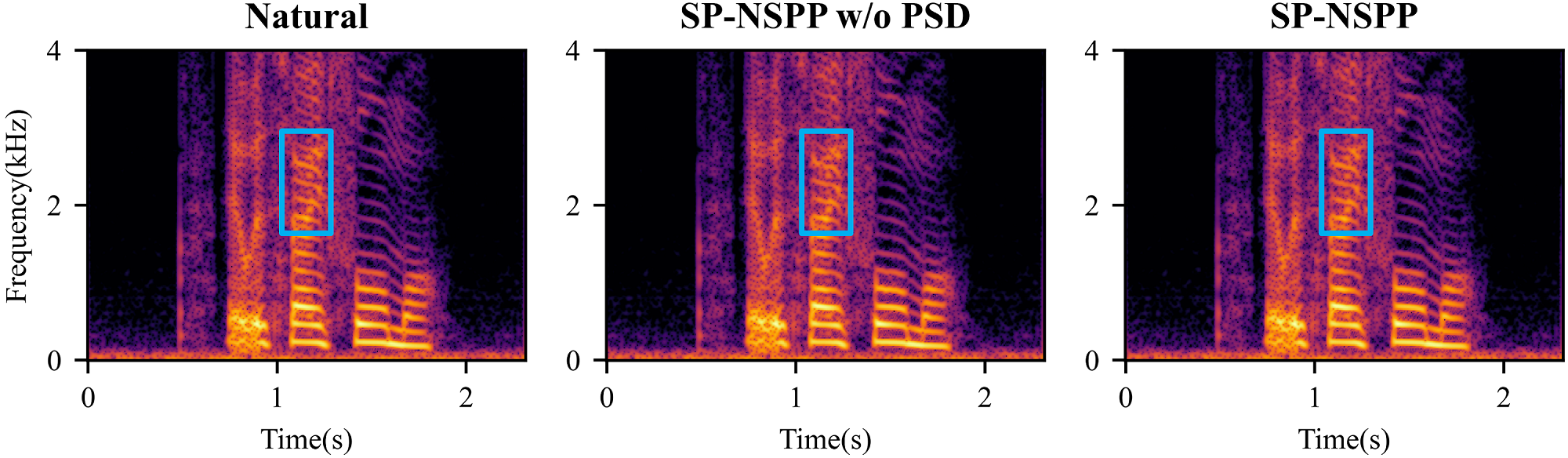}
\caption{A comparison among the spectrograms (0$\sim$4 kHz) of the natural speech and speeches generated by \textbf{SP-NSPP} and \textbf{SP-NSPP w/o PSD} for the analysis-synthesis task.}
\label{fig:spectrum2}
\end{figure}

Then, we conducted three ablation experiments to investigate the roles of key modules in \textbf{SP-NSPP}. 
Three ablated variants were constructed by ablating the refinement stage (denoted by \textbf{SP-NSPP w/o RS}), the PSD (denoted by \textbf{SP-NSPP w/o PSD}), and the TFID loss (denoted by \textbf{SP-NSPP w/o TFID}) from \textbf{SP-NSPP}, respectively. 
The phase accuracy and speech quality evaluation results for the analysis-synthesis task are listed in Table \ref{tab:ablation}. 
It can be observed that all the metrics of \textbf{SP-NSPP w/o RS} lagged behind those of \textbf{SP-NSPP}.
The elimination of the refinement stage had a significant impact on the overall performance of the model. 
This indicates that learning phase patterns directly from amplitude without the guidance of prior phase information is challenging. 
Introducing prior phase information can effectively alleviate the training difficulty of the model, thereby improving the accuracy of phase prediction.
However, the metric results of \textbf{SP-NSPP w/o PSD} are similar to those of \textbf{SP-NSPP}, which may be attributed to the introduction of adversarial training, potentially causing inaccuracies in the objective metrics.
For more evidence, we provided a visual analysis of the spectrograms in Figure \ref{fig:spectrum2}. 
We found that, even after ablating PSD, some harmonic details remained inaccurate (as indicated by the blue box). 
Therefore, the role of PSD is to improve some spectral details and address discontinuities. 
The \textbf{SP-NSPP w/o TFID} significantly lags behind the \textbf{SP-NSPP} in terms of phase continuity metrics (i.e., $\text{PD}_\text{GD}$, $\text{PD}_\text{IAF}$ and $\text{PD}_\text{TFID}$), indicating that the introduced TFID loss effectively enhanced phase continuity.

\subsection{Validation of Generalization under Other Data Conditions}

To further validate the generalization of our proposed \textbf{SP-NSPP} under other data conditions, we compared \textbf{SP-NSPP} with \textbf{RAAR} and \textbf{NSPP} on the VCTK corpus at higher sampling rates (i.e., 24 kHz and 48 kHz) and FSD50K dataset \cite{fonseca2021fsd50k} at 44.1 kHz, which is a human-labeled sound event dataset. 
For the FSD50K dataset, 40,945 utterances and 4,436 utterances were respectively selected as the training and test sets. 

The experimental results for the analysis-synthesis task are presented in Table \ref{tab:48k}. 
For simplicity, we used only three metrics, i.e., $\text{PD}_\text{IP}$, SNR and PESQ.
Under higher sampling rate conditions of speech data, the \textbf{SP-NSPP} still demonstrated impressive performance at 24 kHz, achieving the lowest $\text{PD}_\text{IP}$ and the highest SNR. 
However, at a 48 kHz sampling rate, \textbf{SP-NSPP} was comparable to \textbf{NSPP} in terms of phase accuracy but slightly inferior in speech quality. 
Interestingly, although the iterative algorithm, i.e., \textbf{RAAR}, performed well at a 16 kHz sampling rate as shown in Table 
\ref{tab:main}, its performance deteriorated significantly under high sampling rate conditions of speech data, limiting its applicability. 
In contrast, neural phase prediction models are not limited by changes in sampling rate. 
For non-speech data (i.e., FSD50K), PESQ was removed as it cannot evaluate the perceptual quality of non-speech sounds. 
Our proposed \textbf{SP-NSPP} performed comparably to \textbf{RAAR} and \textbf{NSPP} on non-speech data. 
The above experiments confirm the generalizability of \textbf{SP-NSPP} under different data conditions.


\begin{figure}
\centering
\includegraphics[width=\linewidth]{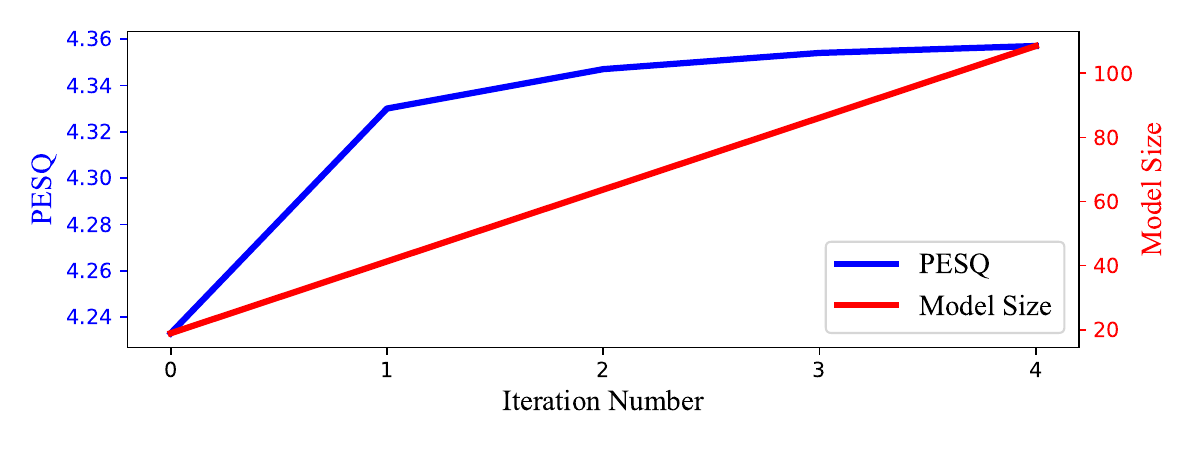}
\caption{Curves of PESQ and model size of the \textbf{SP-NSPP} as a function of the number of iterations for the analysis-synthesis task.}
\label{fig:curve}
\end{figure}

\vspace{+1mm}
\subsection{Discussion on Iterative Prediction}
\label{sec: discussion} 

As mentioned in Section \ref{sec: interative}, the proposed \textbf{SP-NSPP} can also adopt an iterative prediction mode. 
Iteration numbers 0 and 1 correspond to the \textbf{SP-NSPP w/o RS} and \textbf{SP-NSPP}, respectively.
We further increased the number of iterations to 2, 3, and 4, and plotted the curves of PESQ and model size as a function of the number of iterations, as shown in Figure \ref{fig:curve}. 
We can see that as the number of iterations increased, the PESQ also increased, but the growth rate gradually slowed down. 
When the number of iterations increased from 1 to 2, the PESQ rose by less than 0.02, and with further increases in the number of iterations, the PESQ showed almost no significant growth. However, the model size increased linearly with the number of iterations. 
This indicates that PESQ and model size should be balanced. 
An iteration number of 1 (i.e., \textbf{SP-NSPP}), is a good choice as it provides a high PESQ value with moderate model complexity.



\section{CONCLUSION}
\label{sec:con}

This paper presents a novel stage-wise and prior-aware neural speech phase prediction model, named SP-NSPP. 
The prior construction stage generates a prior phase spectrum from the amplitude spectrum, which serves as the conditional input for the subsequent refinement stage. 
With the foundation of the prior phase, the refinement stage can predict a more accurate phase spectrum from the amplitude spectrum. 
To further optimize the phase, we introduce PSD for phase adversarial training and propose the TFID loss which reflects the time-frequency continuity. 
Experimental results demonstrate that our proposed SP-NSPP outperforms traditional iterative estimation algorithms and other neural prediction methods in terms of phase accuracy, speech quality and efficiency for both analysis-synthesis tasks and prediction-synthesis tasks. 
Applying the proposed SP-NSPP to concrete speech generation tasks deeply will be the focus of our future work.


\bibliographystyle{IEEEbib}
\bibliography{strings,refs}

\end{document}